\documentclass[pra,amsmath,twocolumn,showpacs]{revtex4-1}
\usepackage{amsmath,graphicx}
\usepackage{bm}
\usepackage{amssymb}
\usepackage{graphicx,xcolor}
\usepackage{epigraph}
\usepackage{csquotes}
\usepackage{amsmath,amssymb,stmaryrd}
\usepackage{physics}
\usepackage{mathtools}
\usepackage{hyperref}
\hypersetup{
	colorlinks=true,
	linkcolor=blue,
	urlcolor=blue,
	citecolor=blue,
}
\begin{document}
\def\la{{\langle}}
\def\u{\hat U}
\def\A{\mathcal A}
\def\B{\hat B}
\def\C{\hat C}
\def\D{\hat D}
\def\Om{\Omega}
\def\up{\uparrow}
\def\I{\text{Im}}
\def\Re{\text{Re}}
\def\do{\downarrow}
\def\ep{\epsilon}
\def\fb{\overline F}
\def\pia{\hat \pi_A}
\def\wb{\overline W}
\def\nl{\newline}
\def\h{\hat H}
\def\lm{\lambda}
\def\lmu{\underline\lambda}
\def\q{\quad}
\def\t{\tau}
\def\x{\tilde x}
\def\xx{\tilde x_{cl}}
\def\om{\omega}
\def\s{\mathcal{S}}
\def\r{\color{magenta}}
\def\g{\color{teal}}
\def\b{\color{blue}}
\def\ob{\colorbox{orange}}
\def\n{\\ \nonumber}
\def\ra{{\rangle}}
\def\T{{\mathcal{T}}}
\def\Ep{{\mathcal{E}}}
\def\E{{\epsilon}}
\def\a{{\hat a}}
\def\h{\hat{H}}
\def\ha{\hat{H}_A}
\def\ua{\hat U_A}
\def\e{\enquote}
\def\g{\color{teal}}
\def\rr{\color{red}}

\title{Imaginary past of a quantum particle moving on imaginary time }

\date\today
%
%
\author {A. Uranga$^{1,2}$} 
\author {E. Akhmatskaya$^{1,3}$}
\author {D. Sokolovski$^{2,3}$}
\affiliation{$^1$ Basque Center for Applied Mathematics (BCAM), Alameda de Mazarredo 14, 48009, Bilbao, Spain}
\affiliation{$^2$ Departmento de Qu\'imica-F\'isica, Universidad del Pa\' is Vasco, UPV/EHU, Leioa, Spain}
\affiliation{$^3$ IKERBASQUE, Basque Foundation for Science, Plaza Euskadi 5, 48009, Bilbao, Spain}

\begin{abstract}
The analytical continuation of classical equations of motion to complex times suggests that a tunnelling particle spends in the barrier 
an imaginary duration $i|\T|$. Does this mean that it takes a finite time to tunnel, or should tunnelling be seen as an instantaneous process? 
It is well known that examination of the adiabatic limit in a small additional AC field points towards $|\T|$ 
being the time it takes to traverse the barrier. However, this is only  half the story. 
We probe the transmitted particle's history,
  and find that it \textit{remembers} very little of the field's 
past behaviour, as if the transit time were close to zero.
The ensuing contradiction suggests that the question is ill-posed, and we explain why. 
\end{abstract}

%
%
\maketitle
%
%
\section{Introduction}
\vspace{0.1cm}
Recent advances in atto-second laser techniques \cite{AT} have revived the interest in the nearly century-old question \cite{McColl}.
Is tunnelling infinitely fast as was suggested, for example, in references \cite{zero1},\cite{zero2}, or does a particle spend in the barrier a finite duration, as was argued, e.g.,  in \cite{nzero1},\cite{nzero2}?
The question itself is somewhat vague, since there is no consensus as to how this duration should be defined
{(for recent and not-so-recent reviews of the subject see \cite{Rev1}-\cite{Rev2}).}

In dealing with a conceptual difficulty it is often helpful to resort to a simple yet representative model. 
One such model  is that of a semiclassical particle transmitted across a smooth potential barrier. It is well known that the particle's motion can be described by the equations of classical mechanics analytically continued into the complex time plane \cite{Pop}, \cite{McL}. This provides a formal yet not a particularly useful answer: The particle spends in a classically forbidden region an imaginary duration $i|\T|$. The result clearly needs an interpretation, 
and in Ref.\cite{BL} the authors, who studied the adiabatic limit of tunnelling in the presence of a small AC field,  
concluded that $|\T|$ must after all  be 
the time the particle spends in the barrier region \cite{BL1}. In this {paper} we show that analysis of \cite{BL}, \cite{BL1} 
{(the model was recently revisited in \cite{Poll1})}
presents only one side 
of the story. Examining instead the part of the time dependent potential actually experienced by a tunnelling particle 
equally suggests that the duration spent inside the barrier must in fact be close to zero. We argue that the question is ill-posed, 
and explain why it should be so. Before addressing the case of tunnelling we briefly review  the classically allowed case, 
where both approaches agree. 

\section{The classically allowed case}
\begin{figure}[b]
\includegraphics[angle=0,width=8.5cm]{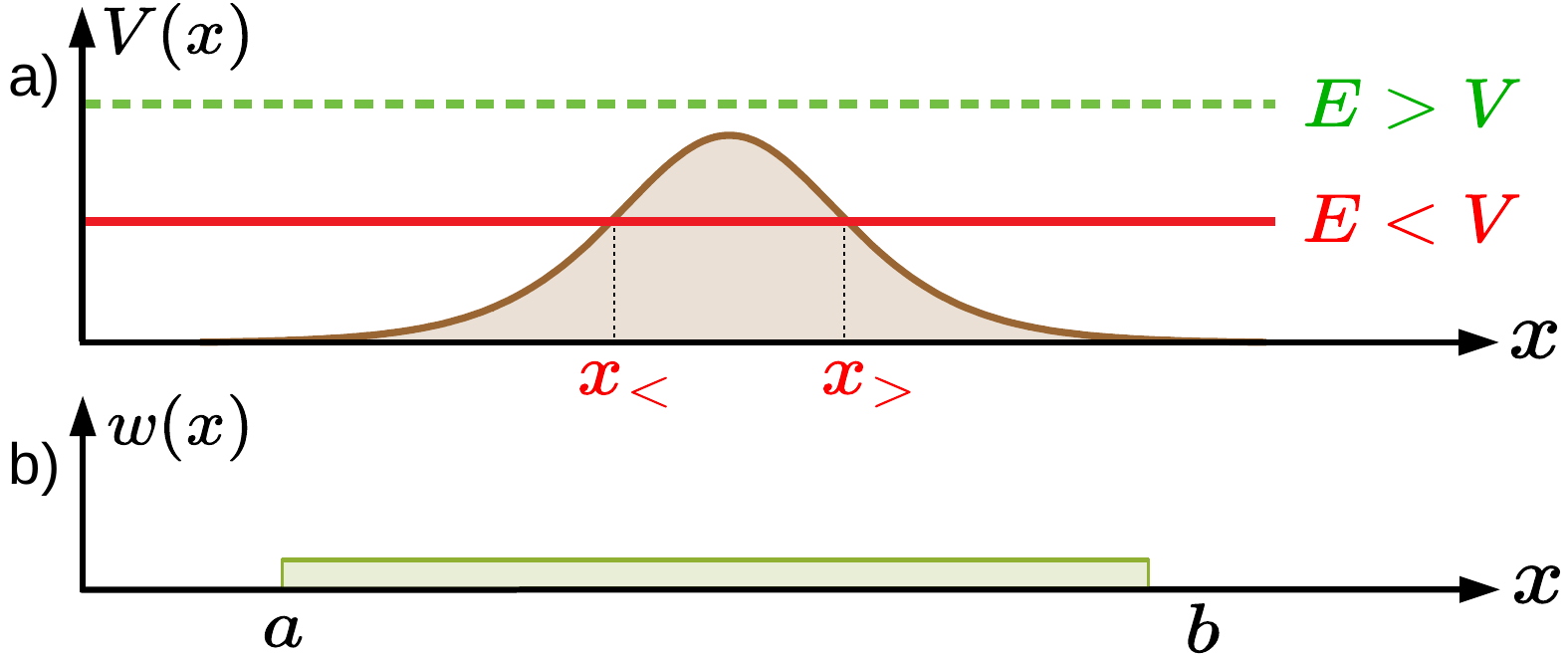}
\caption {a) Depending on the energy $E$, a particle can either tunnel (red, $x_<$ and $x_>$ are the classical turning points), or pass over the barrier's top ({dashed} green).
b) A time-dependent perturbation $W(x,t)=w(x)\Om(t)$ is added between $x=a$ and $x=b$, a constant 
$w(x)=w_0$ is shown. }
\label{Fig.1}
\end{figure}

Consider a particle of energy $E$ and mass $m$, incident from the left on a stationary potential barrier $V(x)$, assuming $E>V(x)$.  A  small time dependent perturbation [$\theta_{ab}(x) =1$ for $a\le x \le b$, and $0$ otherwise]
\begin{eqnarray}\label{0}
W(x,t)=  w(x) \theta_{ab}(x)\Om(t),
\end{eqnarray}
 is added in a region $a< x< b$ (see Fig.\ref{Fig.1}). 
The particle's wave function  $\psi(x,t)$ satisfies a Schr\"odinger equation ($\hbar=1$)
\begin{eqnarray}\label{1}
i\partial_{t}\psi(x,t)=[-\partial_{x}^2/2m +V(x)+W(x,t)]\psi(x,t),
\end{eqnarray}
which can be solved by expanding the wave function in powers of $W$, 
$\psi(x,t)=\sum_{{k}=0}^\infty \psi_{k}(x,t)$,  where
\begin{align}
 \psi_{{k}+1}(x,t)=-i\int_{-\infty}^td\t \int dy K(x,y, t-\t)W(y,\t)\psi_{k}(y,\t),
\end{align}
where  $K(x,y, t-\t)$ is the Feynman propagator \cite{FeynH} for the motion in a potential  $V(x)$.
Evidently, $\psi(x,t)$ may depend on (i.e., ``remember") all the values of $W(y,\t)$, $\t \le t$,
{although its behaviour in the distant past is expected to be less important.
Defining an effective region of integration for an oscillatory integral is notoriously difficult (for more details see, e.g., \cite{DSX1}).
However,} a more precise estimate is available if $V(x)$ varies slowly compared to the particle's de-Broglie wavelength. Then the terms in the perturbation series can be obtained by invoking 
 the semiclassical approximation for $K(x,y,t-\t)$ \cite{FeynH}, 
and evaluating the $\t$-integrals by the stationary phase method.
If, in addition, $W(x,t)$ in Eq.(\ref{1}) also varies slowly in $x$ (except maybe at the endpoints $x=a$ and $x=b$), 
the series can be summed (for details see Appendix \ref{Appx:A}).
 Thus, for an $x>b$, ``downstream" from the region which contains 
$w$, one finds 
\begin{subequations}\label{3}
	\begin{align}
	\psi(x,t)&\approx p(x,E)^{-1/2}\exp\{i[S_0(x,t)+S_1(x,t)]\},\label{3_a}\\
	S_0(x,t)&=\int_a^x p(y,E)dy -iEt,\\
	S_1(x,t)&=-\int_{\t_a}^{\t_b}W(\x(\tau), \tau)d\tau\label{3_c}
	\end{align}
\end{subequations}
and
\begin{eqnarray}\label{a3}
p(x,E)=\sqrt {2m[E-V(x)]}>0, \q\q\q\q\q\q\n
\end{eqnarray}
is the particle's momentum. In Eq. (\ref{3_c}), {$\x(\tau)=\x(\t|x,t,E)$} is the classical trajectory of a particle with energy $E$,  which arrives in  $x$ at a time $t$, 
implicitly defined by
\begin{eqnarray}\label{4}
\tau(\x) = t-\int_{\x}^x \frac{mdy}{p(y,E)}, 
\end{eqnarray}
and {$\t_a\equiv \tau(a,x,t)$,  $\t_b\equiv \tau(b,x,t)$} are the moments  when the trajectory enters and leaves the region  $[a,b]$, respectively.

{Equations (\ref{3})-(\ref{4}), whose validity is discussed in Ref. \cite{DSac} and in Appendix \ref{Appx:B},
 have a simple interpretation. 
We are interested in a particle of energy $E$ found (\e{post-selected}) in a given location $x$ at a time $t$.
In the (semi)classical limit, the particle's past is well defined: it  has been following a classical trajectory $\x(\t|x,t,E)$.
Accordingly, the particle may probe the potential $W$ only at the moment 
$\x(\t|x,t,E)$ passes through that location.}
Clearly it spends inside the region $[a,b]$ a duration 
\begin{eqnarray}\label{b4}
\T_{ab}\equiv \tau_b-\tau_a= \int_a^b \frac{mdy}{p(y,E)}.
\end{eqnarray}
\begin{figure}[h]
\includegraphics[angle=0,width=8.6cm]{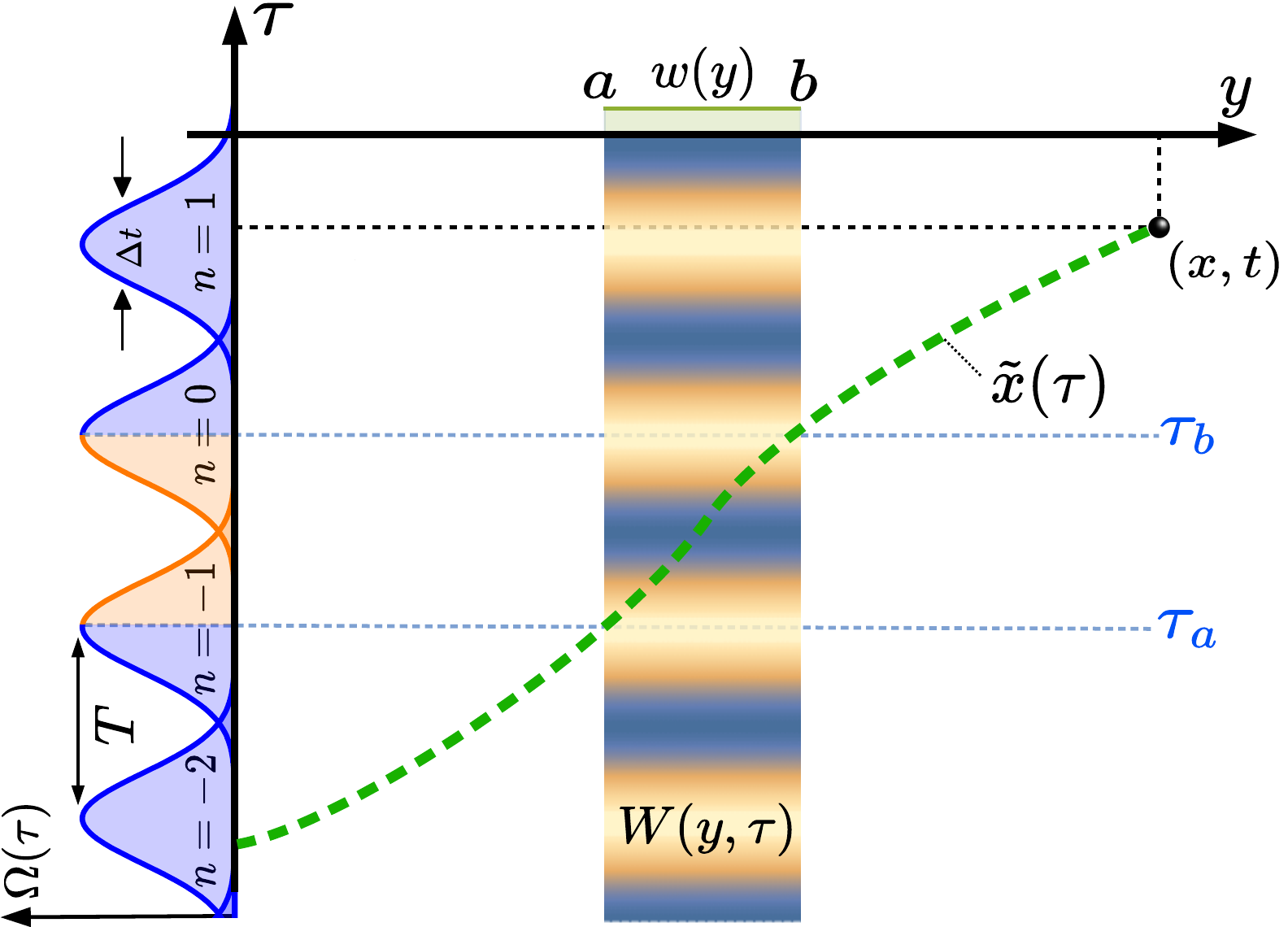}
\caption {The classically allowed case: A particle with energy $ E $ arriving  in $x$ at time $t$ probes the values 
of {$W(\x(\t),\t)$} along its trajectory $\x(\t|x,t,E)$ {(dashed green)} defined by Eq.(\ref{4}). 
Only two pulses, $n=0$ and $n=-1$ {(in orange)}, contribute to the phase of $\psi(x,t)$ (see Table \ref{table:1}).}
\label{Fig.2}
\end{figure}
\newline
It is easy to devise simple tests which would convince one that this is, indeed, the case.
Consider the case of a time dependent potential ``step" 
\begin{eqnarray}\label{6}
W(x,t)=w_0\Omega(t), \q w_0=const. 
\end{eqnarray}
and ask how slowly $\Om$ must vary for the particle to ``see" a static potential, fixed, say,
at the moment it leaves the region $[a,b]$. Choosing $x=b$, $\t_b=t$ helps discount the time of travel from $b$ to $x$,  and expanding  $\Om (\t)$ 
in a Taylor series, $\Om (\t) \approx \Om (\t_b) + \dot{\Om}(\t_b)(\t-\t_b)$,
one finds
\begin{eqnarray}\label{7}
S_1(x,t) \approx w_0\Om(t) \T_{ab} - w_0 \dot{\Omega}(t)\T^2_{ab}/2.
\end{eqnarray}
The first term is the desired adiabatic approximation.
The second one can be neglected if it is small compared to unity. 
 For a harmonic perturbation $\Om(t) = \cos(\om t)$ this means
$\om \T_{ab} \ll  1/ w_0 \T_{ab}$, where $w_0 \T_{ab}\sim 1$,
[cf. Eq.(\ref{B3a})].

As expected, adiabaticity is achieved provided the particle crosses the region $[a,b]$
so quickly, the oscillating potential has no time to change.
 
Our second test is even simpler. Suppose $\Om(t)$ in Eq. (\ref{6}) is a sequence of Gaussian pulses 
of various magnitudes $\beta_n$,
\begin{eqnarray}\label{8}
\Om(\t)\equiv  \sum_{n}{ \Om_n(\t)} = \sum_{n} \beta_n \exp[-(\t-nT)^2/\Delta t^2],\q\q\n
  \Delta t < T, \q {T \le  \T_{ab}},  \q\q\q\q\q\q\q\q\q
\end{eqnarray}
provided by the experimenter. How many of these pulses affect the phase of the wave function ${\psi(x,t)}$ at some $(x >b,t)$?
The answer is, of course, those which overlap with the interval $\tau_a \le \t \le \t_b$, since
\begin{eqnarray}\label{a8}
S_1(x,t)= \sum_{n} S_{1,n}(x,t)=-w_0 \sum^N_{ n=-\infty}\int_{\t_a}^{\t_b} \Om_n(\t)d\t. \q\q
\end{eqnarray}
Figure \ref{Fig.2} shows the case where the particle leaves the region $[a,b]$ at $\t_b=0$, $\T_{ab}$ is equal to the separation 
between the pulses, $T$, and $\Delta t= T/3$. Their relative contributions $|S_{1,n}(x,t)|/|S_{1,0}(x,t)|$ are given 
in Table \ref{table:1} for future comparison with the classically forbidden case.
\begin{table}[]
	\caption{The ratios $|S_{1,n}(x=b,t=0)|/|S_{1,0}(x=b,t=0)|$ for the cases shown 
		{in Fig. \ref{Fig.2} (allowed, $T=\T_{ab}=3\Delta t$) and Fig. \ref{Fig.4}  (forbidden, $T=|\T_{x_{_<}x_{_>}}|=3\Delta t$)}. 
		[Obtained for an Eckart potential $ V(x) $, cf. Ref. \cite{Eck_P}.]}
	\centering
	\begin{ruledtabular}
		\begin{tabular} { r l l }
			{$ \q n $} & \textbf{Allowed} & \textbf{Forbidden} \\
			\colrule
			$ 3 $   & $ 4.1\times 10^{-37} $   & $ 1.5\times 10^{-36} $    \\
			$ 2 $   & $ 2.1\times 10^{-17} $   & $ 9.6\times 10^{-17} $    \\
			$ 1 $   & $ 2.2\times 10^{-5} $    & $ 8.1\times 10^{-5} $     \\
			$ 0 $   & $ 1 $                    & $ 1 $                     \\
			$ -1 $  & $ 1 $   				   & $ 8.1\times 10^{-5} $     \\
			$ -2 $  & $ 2.2\times 10^{-5} $    & $ 9.6\times 10^{-17} $    \\
			$ -3 $  & $ 2.1\times 10^{-17}$    & $ 1.5\times 10^{-36} $    \\
		\end{tabular}
	\end{ruledtabular}
	\label{table:1}
\end{table} 
The same conclusion can also be drawn from inspecting directly observable quantities.
The simplest choice would be the probability density $\rho (x,t) \equiv |\psi(x,t)|^2$, but it remains unchanged by the presence of $W(x,t)$.
The  probability current at $(x,t)$, $j(x,t)= m^{-1}\I[\psi^*(x,t)\partial_x \psi(x,t)]$ may, however, be affected since $W(x,t)$ can alter the particle's velocity.  Thus, for the extra term added to $j_0(x,t)= m^{-1}$
we find
\begin{eqnarray}\label{10}
\delta j(x,t) =\frac{1}{p^2(x,E)}\left [w(b)\Om(\t_b)-w(a)\Om(\t_a)\right ]\n
-\frac{1}{p^2(x,E)}\int_{\tau_a}^{\tau_b}\partial_{\x}[w(\x(\tau))\Om(\tau)]
\dot{\x}(\tau)d\tau. 
\end{eqnarray}
Since $-\partial_{\x}[w(\x(\tau))\Om(\tau)]$ is the additional force acting on the particle,
  the last integral is the work done 
to change its kinetic energy. The first term in brackets accounts for the jolts experienced by the particle as it enters and leaves the 
region, and is the only remaining contribution in the case of  a step-like potential (\ref{6}). The current at $(x,t)$ depends 
on $\Om(\t)$ only for $\t_a <\tau <\tau_b$, which provides yet another proof of the particle's presence between $x=a$ and $x=b$ 
during this interval. 
Note that there is no extra current in the adiabatic limit $\dot \Om \to 0$, since for a static potential the net work done by the extra force acting on the particle is null.
However, for $W$ in Eq. (\ref{6}),  $\delta j(x,t)$ approaches its zero limit as 
$ \sim \dot\Om \T_{ab}$,
and the classical duration spent in $[a,b]$ is again the relevant time parameter.
\par
In summary, both tests are consistent with a classical picture of a particle which enters the region at { $\t=\t_a$, 
leaves at $\t=\t_b$}, and spends there a duration 
$\t_b-\t_a$. 
The  tests can also be applied in the classically forbidden case, and we will do it next.  

\section{The classically forbidden  case (tunnelling)}
\begin{figure}[t]
\includegraphics[angle=0, width=8.8cm]{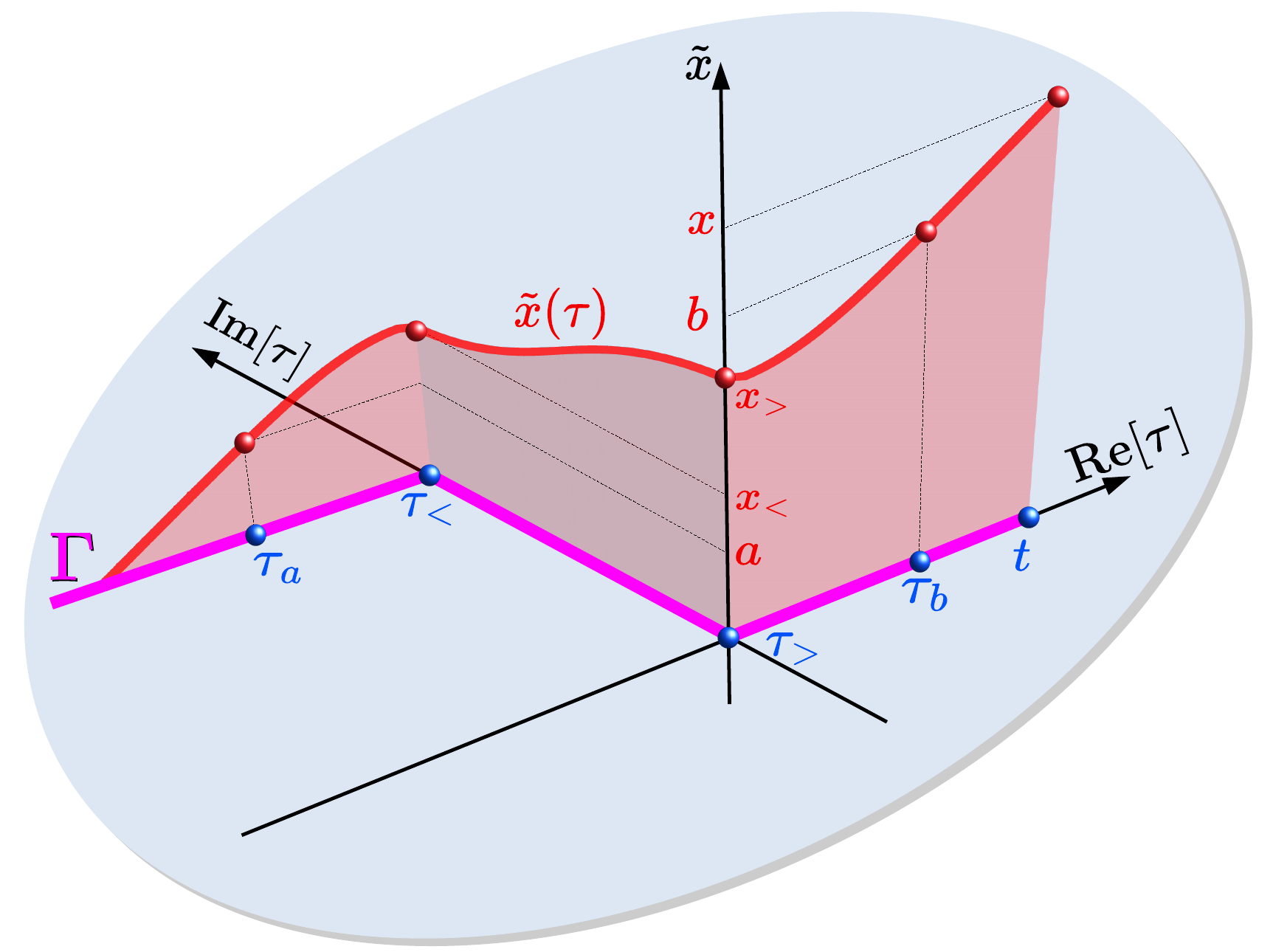}
\caption {The classically forbidden case: A real-valued trajectory {(red)} defined by Eq.(\ref{2b}) along the contour $\Gamma$ {(pink)} in the complex $\tau$-plane.} 
\label{Fig.3}
\end{figure}

The formal derivation in the case  {$E < \max\left[V(x)\right]$} 
is remarkably similar, if one extends the definition of the particle's momentum $p(x,E)$ to the region where $E<V(x)$ as 
($\sqrt x  >0$ for $x>0$)
\begin{eqnarray}\label{0b}
p(x,E) = \begin{cases}
\sqrt{2m[E-V(x)]} &\text{if}\q E\ge V(x)\\
i\sqrt{2m[V(x)-E]} &\text{if}\q E<V(x). 
\end{cases}
\end{eqnarray}
It is easy to check (see Appendices \ref{Appx:A} and \ref{Appx:B}), 
 that the wave function of a particle transmitted across the barrier is still given by Eq.(\ref{3})
where 
\begin{subequations}\label{a0b}
	\begin{align}
	S_0(x,t)&=i \int_{x_<}^{x>}|p(y,E)|dy + \int_{x_>}^x p(y,E) dy -Et,\\
	S_1(x,t)&=-\int_{\t_a\in \Gamma }^{\t_b \in \Gamma}W(\x(\tau), \tau)d\tau,
	\end{align}
\end{subequations}
and $x_<$ and $x_>$ are the turning points where $p(x,E)$ vanishes (see Fig.\ref{Fig.1}).
The main difference with the classically allowed case is that now integration  over $\t$ is performed
 along a contour $\Gamma$ in the complex 
$\t$-plane, defined by, 
\begin{eqnarray}\label{2b}
 \Gamma: \tau(\x) = t-\int_{\x}^x \frac{mdy}{p(y,E)}. 
\end{eqnarray}
The contour, shown in Fig.\ref{Fig.3}, runs along the real $\t$-axis while $\x(\tau) \ge x_>$, then parallel to the imaginary axis 
for $\x$ between $x_<$ and $x_>$, and finally parallel to the real axis for $\x < {x_<}$. 
It is readily seen that on $\Gamma$ Eq. (\ref{2b}) defines a real-valued trajectory $\x(\t|x,t,E)$ ($\I[\x]=0$),
 shown in Fig. \ref{Fig.3} for an Eckart potential \cite{Eck_P}. 
The trajectory satisfies the Newton's equation $m\ddot \x(\tau) = -V'[\x(\tau)]$
along each segment of the contour, and was proven by McLaughlin \cite{McL} to be the real-valued critical path of the Feynman path integral
analytically continued to complex times. The time integrals in the power series for $\psi(x>b,t)$ can now be evaluated by the saddle point method (see Appendix \ref{Appx:A}), and the summed series bears similarity to that in Eq.(\ref{3}).
\par
\begin{figure}[t]
	\includegraphics[angle=0,width=8.6cm]{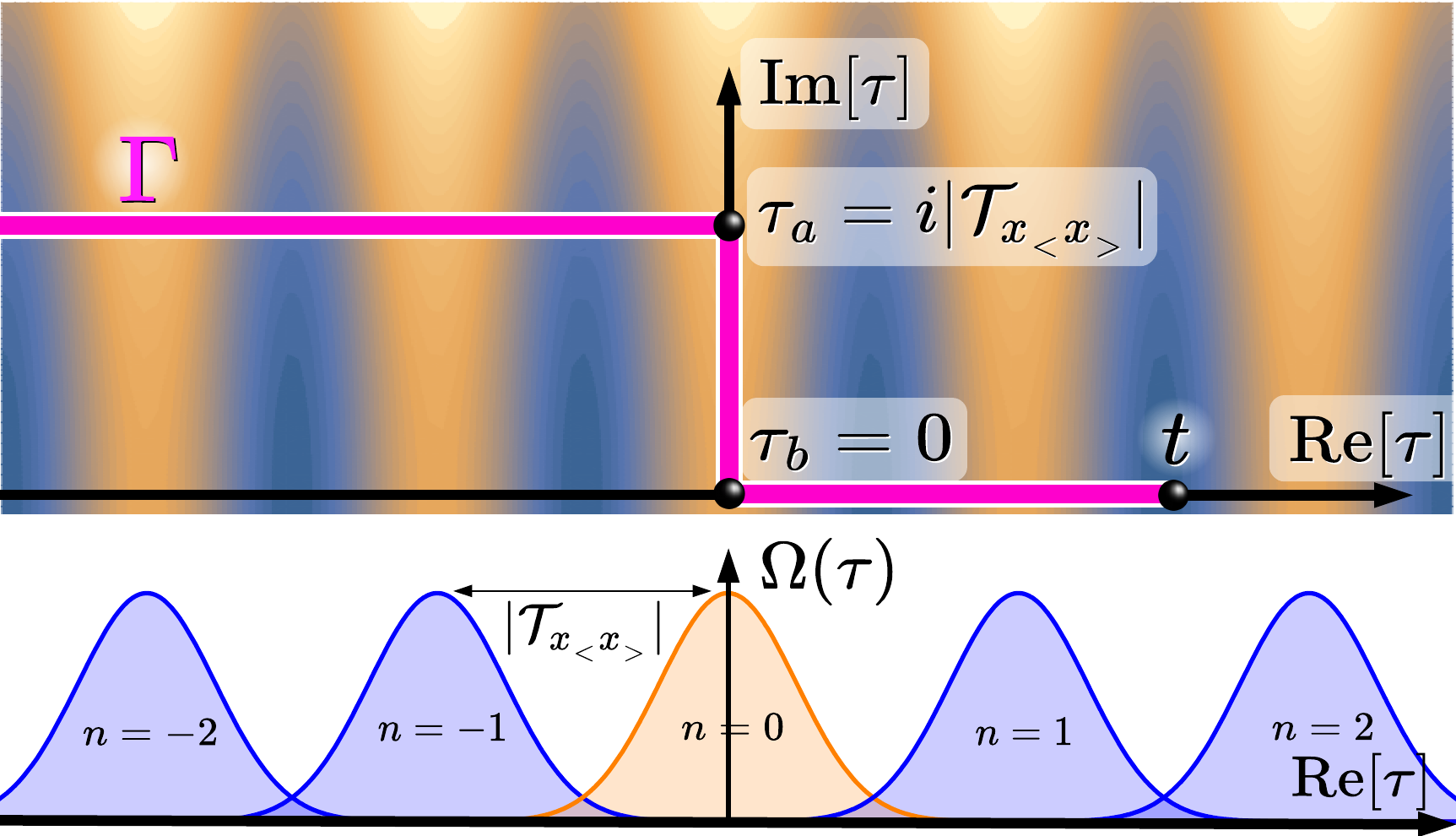}
	\caption {The classically forbidden case: the tunnelling particle probes the $W$ along the vertical part of the contour $\Gamma$. $|\Om( \t)|$
		in Eq. (\ref{8}) is shown as a contour plot. Only the $n=0$ pulse {(in orange)} contributes to $\rho(x,t)$. {($ T= |\T_{x_{_<}x_{_>}}|=3\Delta t$.)}}
	\label{Fig.4}
\end{figure}

There is, however, an apparent difficulty in trying to establish when exactly the tunnelling particle was present in the classically forbidden region,
since a  perturbing potential (\ref{6}), restricted to the sub-barrier region, $a=x_<$, $b=x_>$,  is probed along the vertical part of the contour $\Gamma$ in {Fig. \ref{Fig.4}}.
For a particle, just emerging from the barrier at $x=x_>$,  $t=\t_b={0}$, { $ \t_a=i|\T_{x_{_<}x_{_>}}| $, the integration is along the imaginary time axis, and} one has
 \begin{eqnarray}\label{4b}
S_1(x,t)=iw_0 \int_0^{|\T_{x_{_<}x_{_>}}|}\Om(t+iz)dz,\n
\T_{x_<x_>}=\tau_b-\t_a= -i\int _{x_<}^{x_>} \frac{mdy}{|p(y,E)|}.
\end{eqnarray}
 An experimenter, \textcolor{black}{controlling the real-time behaviour} of the perturbing potential given by (\ref{6}), is unlikely to be satisfied  with an
 explanation that \e{the particle is present inside the barrier between the times $\t_a=i |\T_{x_{_<}x_{_>}}|$ and { $\t_b=t$}}, and would want to know what this could mean  in practice. 
In case of tunnelling, he/she will need to look no further than the particle density at  $(x,t)$ 
\begin{eqnarray}\label{3b}
\rho(x,t)= \rho_0(x) \exp\{ -2\I[S_1(x,t)]\},
\end{eqnarray}
now modified by the perturbation, since $S_1(x,t)$ is complex valued. 
\par
As in our first adiabatic test, one can ask how slowly should $\Om(\t)$ vary on the real $\t$-axis, 
{for $\rho(x,t)$ to be given by its expression for a static step-like potential} {$w_0 \Om$},
in which 
constant {$\Om$} is replaced by $\Om(t)$ for each $t$. Expanding, as before,  $\Om(t+iz)$ in a Taylor series
around {$\t_b$}  yields
 \begin{align}\label{5b}
\I[S_1(x,t)]\approx w_0 \Om(t) |\T_{x_{_<}x_{_>}}|-w_0 \ddot\Om(t) |\T_{x_{_<}x_{_>}}|^3/6.\q
\end{align}
Now with $\Om(t) =\cos(\om t)$ the adiabatic limit $\rho(x,t)=\rho_0(x)\exp[-2w_0\cos(\om t)|\T_{x_{_<}x_{_>}}|]$  
is reached {provided the first and the second terms on the right-hand side of Eq. (\ref{5b}) are of order of unity, 
and much less than unity, respectively.  This leads to a condition
\begin{align}
\om |\T_{x_{_<}x_{_>}}| \ll 1/\sqrt{w_0 |\T_{x_{_<}x_{_>}}|}\sim 1, 
\end{align}
first 
obtained in Ref. \cite{BL}, albeit in a slightly different form.} Its similarity to the classically allowed case 
convinced the authors of Refs. \cite{BL},  \cite{BL1}
 that $|\T_{x_{_<}x_{_>}}|$ 
must be 
``the time a tunnelling particle spends in the barrier region".
We will, however, reserve judgement until submitting the tunnelling particle to our second test. 
\par
Suppose again 
that the potential in the classically forbidden region is modified by a series of Gaussian 
pulses (\ref{8}).  If $|\T_{x_{_<}x_{_>}}|$ is indeed the duration spent in the barrier region, the particle 
exiting the barrier at $x=x_>$ and $t=0$ must be affected
by what happened in the barrier between $t=-|\T_{x_{_<}x_{_>}}|$ and $t=0$.
This, however, does not happen, as we will show next. 
{Calculating $\rho(x,t)$ at $x=x_>$ and $t=0$ with the help of  Eqs.(\ref{4b}) and (\ref{3b})
yields 
\begin{eqnarray}\label{6b}
\I[S_{1,n}(x_>,0)]=w_0\beta_n \exp\left (-\frac{n^2T^2}{\Delta t^2}\right ) \n
\times \int_0^{|\T_{x_{_<}x_{_>}}|} \exp\left (\frac{z^2}{\Delta t^2}\right )
\cos\left( \frac{2nzT}{\Delta t^2}\right ) dz.
\end{eqnarray}
Unlike in the classical case, increasing $|\T_{x_{_<}x_{_>}}|$ (e.g., by making the barrier broader)
does not let the particle experience more pulses, since $\I \left[S_{1,n}\right]/\I\left[S_{1,0}\right]$ remains 
of order of $\exp\left (-\frac{n^2T^2}{\Delta t^2}\right )$, small if $\Delta t \ll T$. 
Moreover, the approximation used in deriving Eqs. (\ref{4b}) and (\ref{3b}) requires
$S_{1,n}\lesssim 1$, so that, by necessity, $\I\left[S_{1,n\ne 0}\right]\ll1$, which leaves $S_{1,0}$ the only non-negligible contribution. The particle appears to have \e{no memory}
of other narrow pulses, and the experimenter would measure practically the same density $\rho(x_>,0)$  by applying only}
{the $\Om_{n=0}(\t)$ pulse}.
{The ratios  $|S_{1,n}(x_>,0)|/|S_{1,0}(x_>,0)|$
for 
\begin{align}\label{6ba}
T=|\T_{x_{_<}x_{_>}}|, \q \Delta t =T/3,
\end{align}
are listed in Table \ref{table:1} alongside the  results obtained earlier for the classically allowed case. 
With our choice of parameters,  in the classical regime the pulses with $n=0$ and $n=-1$ 
contribute equally to $ S_1 $ in the exponent in Eq.(\ref{3_a}). With tunnelling, only the $n=0$ pulse affects the particle emerging from the barrier, 
which should not be the case for a particle spending $|\T_{x_{_<}x_{_>}}|$ in the sub-barrier region.} 

There is obviously a discrepancy with what was learnt in the adiabatic test, and it is easy to see why. In our first test, the rate of change of a slowly varying analytical function 
$\Om(\t)$ along any direction in the complex $\t$-plane is determined by the same $\dot\Om(\t_b)$,
which accounts for the similarity between Eqs.(\ref{7}) and (\ref{5b}). 
However, in the second test, analytic continuation of a Gaussian is such that 
making  {barrier} {\it broader}  does not let the particle explore its behaviour further into the past. 
\textcolor{black}{(}The reader may note a certain parallel with the Hartman effect \cite{Hart}, whereby the arrival of a tunnelled particle 
 is not delayed by increasing the barrier's width.)
The effect persists if the ratio $\Delta t /|\T_{x_{_<}x_{_>}}|$ is made smaller (see Appendix \ref{Appx:B}), 
which suggests that the duration the particle spends in the barrier must be zero, or very close to zero. 
{One can keep the ratio $\Delta t /T$ constant, and reduce $T$ so more pulses fit into the interval 
$\left[-\left|\T_{x_{_<}x_{_>}}\right|,0\right]$  ($w_0$ can be chosen to ensure $S_{1}\sim 1$.) Still, only the $n=0$ pulse will be taken into account, even though now the $n=-1$ pulse belongs to a more recent past.}

\section{Conclusions and discussion}
{Our aim here is not to provide yet another \e{evidence} of tunnelling being \e{instantaneous}, but rather to gain a broader view of the problem itself. 
Two equally feasible (at least in principle)  experiments give the same answer in the classically allowed case. 
Both suggest that the interaction of a classical particle with a small field is governed by the duration $\T$, spent by its trajectory region where the field is applied. 
Alice and Bob, making different tests in their respective laboratories, can agree about this point. 
\newline
This is not so in the case of tunnelling. An extension of the classical equations of motion to complex time still yields a
single \e{duration} $\T$, but now it is imaginary, $\T=i|\T|$. Is tunnelling instantaneous, since $\Re[\T]=0$, 
or does it take $|\T|$ seconds to cross the barrier region?
Both tests can be performed also in the tunnelling regime, but this time the results appear to contradict each other.
With the classical picture still in mind, Alice's would find her adiabatic test pointing towards a finite tunnelling time, $|\T|$. This is the well-known
B\"uttiker-Landauer result, often interpreted as the true duration of  a tunnelling process \cite{BL}. On the other hand Bob, who studies the 
\e{memory} of  a tunnelled particle, must conclude that the duration spent in the barrier is much shorter than 
$|\T|$, and is in fact close to zero. 
If both Alice's in Bob's results are correct, the concept of a well-defined classical-like tunnelling time must be at fault.
\newline
The problem is well known in quantum measurement theory.
 A quantity, uniquely defined on a single (classical) trajectory must become indeterminate if two or more paths leading to the 
particle's final condition interfere \cite{FeynL}. An attempt to measure it without destroying the interference yields only a complex \e{weak value}  (see, e.g., Ref. \cite{DSwv} and references therein).
Such is the time measured by a weakly coupled Larmor clock \cite{Larm1}-\cite{Larm2}, $\t^L=\t^L_1+i\t^L_2$.
The spin of the clock results rotated both in the plane perpendicular to the magnetic  field by 
an angle $\varphi_1\sim \t_1$ (as in the classical case), but also in the plane parallel to the field, by $\varphi_2\sim \t_2$ (a new non-classical feature).
Moreover, in the semiclassical limit one has $\t_1 \approx 0$, whereas $\t_2 \approx |\T|$ (the reader may begin to see a connection). 
Alice, who has measured $\varphi_1\approx 0$, could argue that \e{tunnelling is instantaneous}, since the spin  has not rotated in the way it does in the classical case. Bob, who has measured $\varphi_2\approx |\T|$, could argue that tunnelling \e{takes a finite time} since, after all, the field caused the spin rotate, albeit in a different manner. So, which part of the complex $\t^L$ represents the physical \e{duration spent in the barrier}? 
\par 
In general, neither $\t_1$ nor $\t_2$ is a valid candidate since, for some initial and final particle's states, both can turn out to be negative \cite{DSsw}, \cite{DSel}. 
The absolute value, $|\t^L|$ (known as the \e{B\"uttiker time} \cite{BL2}) is indeed real, but can exceed the time the particle was in motion \cite{DSsw}, \cite{DSel}. The reason for this strange behaviour is the Uncertainty Principle \cite{FeynL} which, among other things, forbids knowing the time spent in the barrier in a situation where different durations interfere to produce the tunnelling amplitude \cite{DSel}. 
 \par
If the theory does not allow one to describe classically forbidden tunnelling in terms of a classical-like duration, it is rather pointless to ask whether or not this non-existent duration should be zero. One can try to approach the question from the operational side, by adopting to the tunnelling regime experiments, known to yield the classical duration in the classically allowed case. The problem is, one will obtain results which, if interpreted
at face value, appear to support either of the conflicting viewpoints. In this paper we have shown that the same is also true for the 
B\"uttiker-Landauer model  \cite{BL}, which employs a time-modulated barrier, instead of an external clock. 
A reasonable  way out of the contradiction is to accept that, in the absence of a classical-like  tunnelling time, different experiments
may yield different quantities with units of time \cite{Poll2}.  The nature and properties of such quantities can then be analysed 
individually by the methods of elementary quantum mechanics.
\par
One may be disappointed to learn that the object of a long standing search may not exist or, what is worse,
exists in many conflicting shapes and forms. 
Perhaps a different approach would deliver a well defined tunnelling time? This seems unlikely. The \e{phase time} (see, e.g., Ref. \cite{Rev1}), associated with the transmission of broad wave packets, can be analysed along the same lines as the Larmor time \cite{DSX1}, \cite{DSno}, \cite{DSX2}. Moreover, the above analysis emphasises 
the importance of a conventional trajectory for the existence of a unique duration spent in a given region of space.
Even when such a trajectory exists along a contour displaced into the complex time plane (cf. Fig. \ref{Fig.3}), the resulting complex duration
looses the properties of a physical time interval. }

 \section*{Acknowledgements}
D.S. acknowledges financial support by Grant No. PID2021-126273NB-I00 funded by MICINN/AEI/10.13039/501100011033 and by ``{ERDF A way of making Europe}", as well as by the Basque Government Grant No. IT1470-22.

A.U. and E.A. acknowledge the financial support by the Ministerio de Ciencia y Innovaci\'on (MICINN, AEI) of the Spanish Government through BCAM Severo Ochoa accreditation CEX2021-001142-S and PID2019-104927GB-C22, PID2022-136585NB-C22 grants funded by MICIU/AEI/10.13039/501100011033, as well as by the Basque Government through 
the BERC 2022-2025 Program, IKUR Program, ELKARTEK Programme (Grants No. KK-2022/00006 and No. KK-2023/00016).

\appendix
\section{Quantum perturbation theory}\label{Appx:A}
We need to solve, in the semiclassical limit, a Schr\"odinger equation 
\begin{eqnarray}\label{A1}
\left[i\hbar \partial_t +\hbar^2\partial_x^2/2m -V(x)\right]\psi(x,t)= W(x,t)\psi,
\end{eqnarray}
where $W(x,t)$ vanishes unless $a \le x \le b$.
 \subsection{The classically allowed case}
Expanding $\psi(x,t)$ in a Born series, $\psi=\sum_0^\infty \psi_{k}$, one has
\begin{eqnarray}\label{A2}
{\psi_{k}(x,t) = \int_a^b dy \int_{-\infty}^t d\t }\times\q\q\q\q\q\q\q \n
G(x,y,t-\t)W(y,\t)\psi_{{k}-1}(y,\t), \q {k}=1,2...
\end{eqnarray}
where 
{\begin{align}
\psi_0(x,t)=p(x)^{-1/2}\exp[ i\int^x p(y)dy/\hbar- iEt/\hbar],
\end{align}}
and   $p(x)=\sqrt {2m[E-V(x)]}$.
The Green's function $G(x,y,t-\t)$ is simply related to the Feynman propagator \cite{FeynH}
 $K(x,y,t-\t)$ as
{\begin{align}
G(x,y,t-\t)=(-i/\hbar) K(x,y,t-\t){ \Theta(t-\t)}, 
\end{align}}
where $\Theta(x)$ is $1$ for $x>0$, and $0$  otherwise.
 The propagator has a well known classical limit \cite{FeynH}, 
$K(x,y,t-\t)\sim \exp[i S(x,y,t-\t)/ \hbar]$ where 
{\begin{align}
S(x,y,t-\t) =\int_{\t}^t [m\dot \x^2 (\t')/2-V(\x(\t'))]d\t'
\end{align}}
is the classical action evaluated along the trajectory $\x(\t')$ connecting the points $(y,\t)$ and $(x,t)$.

As  $\hbar \to 0$, the time integral for ${k}=1$ can be evaluated by the stationary phase method. 
The corresponding condition $\partial_{\t}S(x,y,t-\t)-E=0$ ensures that the main contribution to the integral comes 
from a time $\tau_s$ at which a particle with energy $E$, destined to arrive at $(x,t)$, passes through the point $y$, 
\begin{eqnarray}\label{A3}
\t_s(x,y,t,E)=t-\int^x_{y} \frac{m}{p( z)} dz.
\end{eqnarray}
With a pre-exponential factor $W(y,\t)$ evaluated at $\t=\t_s$, the first term of the series \textcolor{black}{[see Eq.(\ref{A2})]} takes the form
\begin{align}\label{A4}
\psi_1(x,t) =(-i/\hbar)\psi_0(x,t) \int^x_{-\infty}dy\frac{m}{ p(y)}W\left (y,\t_s\right).
\end{align}
Repeating the calculation, using a relation 
{\begin{align}
\int_a^b &dx_n \int _a^{x_n}dx_{n-1}...\int_a^{x_2} dx_1
f(x_n)f(x_{n-1})...f(x_1) \nonumber\\
&=\frac{1}{n!}\left [\int_a^b f(x') dx'\right]^n,
\end{align}} 
and summing the series for $\psi(x,t)$ yields 
\begin{align}\label{A5}
\psi(x,t) =\sum_{k=0}^\infty\psi_{k}(x,t)\approx \psi_0 (x,t)\exp [iS_1(x,t)/\hbar],
\end{align}
where, explicitly, 
\begin{align}\label{A6}
S_1(x,t)\equiv -\int^b_{a}dy\frac{m}{ p(y)}W\left (y,t-\int^x_{y}{  \frac{m}{p(y')}} dy'\right). 
\end{align}
Changing the variable $y\to \t(y)  =t-\int^x_{y} \frac{mdy'}{p(y')}$  yields the last of Eqs.(\ref{3}).
 A classical particle ``probes" a time dependent perturbation as it follows its trajectory. 
 \subsection{The classically forbidden case}
The case where a particle tunnels across a barrier $V(x)$ can be treated in a similar manner
(at least in one spatial dimension), as was shown by McLaughlin in Ref. \cite{McL}. After applying the standard connection formulae
  and multiplying the result by $\exp(-i\pi/4)$ the zero-order wave function can be written as  
\begin{widetext}
\begin{equation}\label{A7}
\psi_0(x,t) =
\begin{cases}
p(x)^{-1/2}\Big(\exp[i\int_{x_<}^x p(y)dy-iEt] 
-i \exp[-i\int_{x_<}^x p(y)dy-iEt]\Big), & x <x_<,\\
p(x)^{-1/2}\exp[i\int_{x_<}^x p(y)dy-iEt], & x>x_<
\end{cases}
\end{equation}
\end{widetext}

where  $p(x)=i|p(x)|$  between turning points $x_<$ and $x_>$ and $p(x)=|p(x)|$ otherwise.
As was shown in Ref. \cite{McL}, the semiclassical limit of $K(x,y,t-\t)$ analytically continued into 
the complex plane of $\t$ still given by $K(x,y,t-\t)\sim \exp[i S(x,y,t-\t)/\hbar]$, with
$S(x,y,t-\t)$ 
 evaluated along the real valued path along a complex contour 
connecting complex $\t$ with a real $t$. 
The integrals (\ref{A2})  can now be evaluated by the saddle point method.
For $y < x_<$ the complex saddle at [cf. Eq. (3.9) of Ref. \cite{McL}]
\begin{eqnarray}\label{A8}
\t(x,y,t,E)=t-\int^{x_<}_{y} \frac{m}{p(y')} dy'+\n
i\int^{x_>}_{x_<} \frac{m}{|p(y')|} dy'-\int^{x}_{x_>} \frac{m}{p(y')} dy'
\end{eqnarray}
 lies in the upper half of the complex $\t$-plane.
 For an $y$ lying between the turning  points $x_<$ and $x_>$ (see Fig.\ref{Fig.1}) one has
\begin{align}\label{A9}
\t(x,y,t,E)=t+i\int^{x_>}_{y} \frac{m}{|p(y')|} dy'-\int^{x}_{x_>} \frac{m}{p(y')}dy'
\end{align}
and for $x>x_>$ there is real stationary point 
\begin{eqnarray}\label{A10}
\t(x,y,t,E)=t-\int^{x}_{y} \frac{m}{p(y')} dy'.
\end{eqnarray}
 The only difference from McLaughlin's derivation \cite{McL} is the presence of $W(y,\t)$ in the integrand, which needs to be evaluated at the saddle. Acting as before, we obtain Eq.(\ref{A5})
 with $\psi_0(x,t)$ given by the last of Eqs.(\ref{A7}) and a complex valued 
 \begin{align}\label{A11}
S_1(x,t)=  -\int^b_{a}dx'\frac{m}{ p(x')}W\left (x',\t(x,y,t,E)\right ),
\end{align}
which after a change of variables becomes Eq.(\ref{a3}), where it is understood that the ``time" $\t$ varies along the complex contour
$\Gamma$, defined by Eq.(\ref{2b}).
 A tunnelling particle ``moving on complex time along a real valued trajectory" \cite{McL} ``probes" a time dependent perturbation 
 analytically continued away from the real time axis. 
 \section{Classical perturbation theory}\label{Appx:B}
 The result (\ref{A5})
 can be obtained in a different and, perhaps, a more illustrative manner.
\subsection{The classically allowed case}
 Making the usual ansatz (and restoring $\hbar=1$) $\psi(x,t)= A(x,t) \exp[i{S(x,t)}]$,
one finds that $S(x,t)$ satisfies the Hamilton-Jacobi equation, 
 \begin{align}\label{B1}
\partial_t S(x,t)+[ \partial_xS(x,t)]^2/2m +V(x)+W(x,t)\theta_{ab}(x) =0,
\end{align}
 where $\theta_{ab}(x)=1$ for $a\le x \le b$ and $0$ otherwise.
If $ W $ is sufficiently small, it can be neglected in the pre-exponential factor $A\textcolor{black}{(x,t)}$,
 but not in the exponent,  where one writes $ S(x,t)=\sum_{{k}=0}^\infty S_{k}(x,t)$.
 Equation (\ref{B1}) can be solved iteratively.  Setting
$S_0(x,t)=\int_{a}^x p(y)dy-Et$ 
yields  a first order equation for $S_1$, 
\begin{eqnarray}\label{B2}
\partial_t S_1(x,t)+ p(x)\partial_x S_1(x,t)/m+W(x,t)\theta_{ab}(x)=0,\q\q\q
\end{eqnarray}
which has a particular solution
\begin{eqnarray}\label{B3}
S_1(x\ge a,t) =-\int_a^{\min(x,b)}\frac{mdy}{p(y)}W\left(y,t-\int_{y}^x\frac{mdy'}{p(y')}\right),\q\q\n
\end{eqnarray}
$S_1(x< a,t)\equiv 0$.
The higher-order  terms,  $S_{k}$ with ${k}>1$,  can be neglected if they are small compared to unity. This requires  
 \begin{eqnarray}\label{B3a}
 S_0(x,t) \gg\hbar , \q S_1(x,t) \sim \hbar,\q S_{{k}>1}(x,t) \ll \hbar,  \q
 \end{eqnarray}
which, in turn, imposes restrictions on the magnitude and the rate of change of a perturbation $W(x,t)= {w(x)}\Om(t)$. 
\newline 
Setting $\Om=const$ and applying  (\ref{B3}) for $a<x<b$ yields
\textcolor{black}{$\psi(x,t) \approx p(x,V)^{-1/2}\exp\{i\int_a^x \left [p(y,V)
-\frac{mW(y)}{p(y,V)}\right ]dy-iEt\}$}.
This agrees with the correct semiclassical form \textcolor{black}{[$p(x,V+W)\equiv\sqrt{2m(E-V(x) -W(x)}$]},
\textcolor{black}{$\psi(x,t) = p(x,V+W)^{-1/2}\exp\{i\int_a^x p(y,V+W)dy-iEt\}$} provided
 \begin{eqnarray}\label{B4}
W \ll  E-V \sim p^2/2m. 
\end{eqnarray}
\newline
The limit on how fast  $W(x,t)$ can change with time for the approximation to remain valid is obtained by 
considering the first order term of the Born series. Writing $\Om(t) = \int d\om \Om(\om) \exp(i\om t)$ one notes  that   
after absorbing (emitting) quanta $\psi(x,t)$ for $x>b$ should contain terms like  {$p(E+\om)^{-1/2}\exp[i\int^x p(y,E+\om)dy-i(E +\om)t]$}.
This agrees with 
{\begin{align}
p(E)^{-1/2}\exp[i\int^x p(y,E)dy+i\om\int^x \frac{mdy}{p(y,E)}-i(E +\om)t]
\end{align}}
obtained from Eqs.(\ref{3}) and (\ref{B3}) provided the absorbed energy is small compared to the particle's kinetic energy, $\om \ll  E-V$. If  $\Delta t$ is the time scale upon which $W(x,t)$ changes, 
the condition reads 
 \begin{eqnarray}\label{B5}
1/\Delta t  \ll  E-V \sim p^2/2m. 
\end{eqnarray}
\newline
Finally, replacing in Eq.(\ref{B3}) $y\to \t(y)=t-\int_{y}^x\frac{mdy'}{p(y')}$, for $x>b$ 
one recovers Eq.(\ref{a3}). 
 \subsection{The classically forbidden case}
$S_1(x,t)$ in Eq.(\ref{B3}) remains a continuous solution of Eq.(\ref{B2}) if $p$ is defined by Eq.(\ref{0b}). 
Thus,  $\psi(x,t)$ in Eq.(\ref{3}) is valid, provided the presence of $W(x,t)$ does not alter the matching rules
near the turning points $x=x_<$ and $x=x_>$, where conditions (\ref{B4}) and (\ref{B5}) are clearly violated. 
However, a more detailed analysis (see Ref. \cite{DSac} and references therein) shows that the rules remain unchanged
near a linear turning point, $V(x) \approx \partial_x V(x_>) (x-x_>)$, $W(x,t) \approx W(x_>,t)$. 

The simplest illustration is given by the case of a broad rectangular barrier, $V(x)=V\theta_{ab}(x)$, 
where the matching rules at $x=b$ depend on the particle's energy $E$ (see Ref. \cite{DSac}).
Condition (\ref{B5}), applied on both sides of the turning point, eliminates this dependence, leaving $\psi(x>b,t)$
 still given by Eq. (\ref{3}), apart from an overall energy-dependent factor. 
 Suppose, for simplicity, that $E=V/2$, and $\Om(t)=\exp(-t^2/\Delta t^2)$.
We are interested in the ratio between the pulse's width $\Delta t$ and the modulus of the complex time,
$|\T_{x_{_<}x_{_>}}|=|\T_{ab}|=m(b-a)/\sqrt{mV}$. A typical absorbed energy is obviously of order of
$\om \sim 1/\Delta t$,  and from (\ref{B5}) we have
 \begin{eqnarray}\label{B6}
\Delta t/|\T_{ab}| \gg  2/|p|(b-a).
\end{eqnarray}
However, $|p|(b-a)\gg 1$ is the sub-barrier action, which is always large in the semiclassical limit considered here. 
This shows that the used approximation (\ref{3}) is valid if the pulse is significantly shorter than the B\"uttiker-Landauer time $|\T_{ab}|$ \cite{BL}.


\end{document}